\documentclass[
aps,
prd,
reprint,
superscriptaddress,
footinbib,
nolongbibliography,
floatfix
]{revtex4-2}

\usepackage{amsmath,amssymb,bm,mathtools}
\usepackage{xcolor}
\usepackage[colorlinks=true,citecolor=blue,urlcolor=blue,linkcolor=blue]{hyperref}
\usepackage{microtype}
\usepackage{CJK}
\usepackage{tikz}
\usepackage{orcidlink}
\usetikzlibrary{arrows.meta,calc,decorations.pathmorphing}

\newcommand{\shuffle}{\mathbin{\amalg}}
\newcommand{\dd}{\mathrm d}
\newcommand{\PT}{\operatorname{PT}}
\newcommand{\KN}{\operatorname{KN}}
\newcommand{\pos}{\operatorname{pos}}
\newcommand{\Tr}{\operatorname{Tr}}

\newcommand{\cP}{\mathcal P}
\newcommand{\bF}{\mathbf F}
\newcommand{\bJ}{\mathbf J}

\newcommand{\Part}{\operatorname{Part}}
\newcommand{\Hplus}{\mathbb H_+}

\tikzset{ffscalar/.style={line width=.45pt},
ffgluon/.style={decorate,decoration={snake,amplitude=.6pt,segment length=3pt},line width=.45pt}}
\newcommand{\ffexchange}[3]{\mathord{\begin{tikzpicture}[x=.70cm,y=.70cm,baseline=-2pt,every node/.style={font=\scriptsize,inner sep=1pt}]
\draw[#1] (0,0)--(.7,0);
\draw[#1] (.7,0)--(1.8,0);
\draw[#1] (1.8,0)--(2.5,0);
\draw[line width=.4pt] (0.6649999999999999,.04)--(0.6649999999999999,.32);
\draw[line width=.4pt] (0.735,.04)--(0.735,.32);
\fill (.7,0) circle (.7pt);
\draw[line width=.4pt] (1.7650000000000001,.04)--(1.7650000000000001,.32);
\draw[line width=.4pt] (1.835,.04)--(1.835,.32);
\fill (1.8,0) circle (.7pt);
\node[above] at (.7,.32) {$q_1$};\node[above] at (1.8,.32) {$q_2$};
\node[below] at (0,-.03) {$p_{#2}$};\node[below] at (2.5,-.03) {$p_{#3}$};
\end{tikzpicture}}}
\newcommand{\ffcontact}[1]{\mathord{\begin{tikzpicture}[x=.70cm,y=.70cm,baseline=-2pt,every node/.style={font=\scriptsize,inner sep=1pt}]
\draw[#1] (0,0)--(.9,0);\draw[#1] (.9,0)--(1.8,0);
\draw[line width=.4pt] (0.865,.04)--(0.865,.32);
\draw[line width=.4pt] (0.935,.04)--(0.935,.32);
\fill (.9,0) circle (.7pt);
\node[above] at (.9,.32) {$Q$};
\node[below] at (0,-.03) {$p_1$};\node[below] at (1.8,-.03) {$p_2$};
\end{tikzpicture}}}

\newif\ifmarknewrefs
\marknewrefstrue

\begin{document}
\begin{CJK*}{UTF8}{gbsn}	
	\title{General Form Factors from String Amplitudes}
	
\author{Qu Cao (曹趣)\,\orcidlink{0000-0002-0381-9332}}
\email{caoqu@westlake.edu.cn}
\author{Liang Zhang (张亮)\,\orcidlink{0000-0003-2811-1768}}
\email{zhangliang@westlake.edu.cn}
\affiliation{Institute of Natural Sciences, Westlake Institute for Advanced Study, Hangzhou 310024, China}
\affiliation{Department of Physics, School of Science, Westlake University, Hangzhou 310030, China}
\date{\today}
	
\begin{abstract}
We extend the string-amplitude construction of single-insertion form factors to multiple operator insertions carrying independent off-shell momenta. By deforming the bulk closed-string punctures onto the open-string boundary, we establish a relation between stringy general form factors and scattering amplitudes. Using this relation, we evaluate the field-theory limit and find that the resulting form factors receive contributions from both separate operator insertions and local contact terms. Furthermore, we uncover several hidden structures of stringy general form factors at finite $\alpha'$:
$2$-splits and hidden zeros analogous to those of scattering amplitudes, bulk zeros specific to form factors, and a decoupling split unique to multiple insertions.
\end{abstract}
\maketitle
\end{CJK*}
\section{Introduction}
\label{sec:introduction}
Form factors connect on-shell scattering states with operator insertions. On-shell methods and related amplitude techniques provide efficient tools for their computation \cite{Brandhuber:2011ff,Bork:2011ff,Brandhuber:2011harmony,Brandhuber:2012analyticff,Brandhuber:2014remainders,Boels:2013ffCK,Bork:2014polytopes,Frassek:2016onshellff,BorkOnishchenko:2016lightlikeFF,BorkOnishchenko:2017WilsonLines,Yang:2016fiveLoopFF,BorkOnishchenko:2018ambitwistorFF,Bianchi:2019loopRecursion}. Relations to Yang--Mills--scalar (YMS) amplitudes and double-copy constructions provide further connections with scattering amplitudes \cite{Dong:2022relations,LinYang:2022FFdoublecopy,LinYang:2024foundations,LinYang:2024generalizations} \footnote{For reviews, see Refs.~\cite{NandanYang:2018hidden,Yang:2020review}.}.

A complementary string-inspired approach starts from disk amplitudes involving open-string states on the boundary and closed-string insertions in the bulk \cite{GarousiMyers:1996,StiebergerTaylor:2015}. Contour deformation expresses the bulk integrations as a sum of ordered boundary integrals \cite{Stieberger:2009hq}, which obey monodromy relations \cite{Plahte:1970,BjerrumBohr:2009monodromy}. These integrals encode gauge and scalar amplitudes and their string corrections \cite{Mafra:2013disk,CarrascoMafraSchlotterer:2017,HeTengZhang:2019string}. An analytic specialization of the disk invariants recently gave a string-amplitude construction of single-insertion form factors, with the operator momentum assigned to one bulk puncture \cite{Cao:2025ff}.

In field theory, a matrix element of a time-ordered product of multiple operators nsertions defines a \emph{general form factor}. General form factors have been studied through on-shell source constructions \cite{EngelundRoiban:2013multi}, at strong coupling \cite{GaoYang:2013multi}, in perturbative loop calculations \cite{Ahmed:2020double}, and through the double copy \cite{LiYangZhu:2026multi,Li:2026zqu}. With several bulk punctures, does the field-theory limit of the disk integral reproduce only the general form factor for separate insertions, or does it also contain local contact terms?

In this paper, we extend the string-inspired construction of single-insertion form factors to multiple insertions carrying independent off-shell operator momenta. We refer to the resulting objects as \textit{stringy general form factors}, and derive their field-theory limit. This limit is a sum of contributions from separate operator insertions and local contact terms, rather than a single form factor with all insertions kept separate. 

At finite $\alpha'$, the resulting string integrals extend the $2$-splits and hidden zeros of tree amplitudes \cite{ArkaniHamed:2023zero,Cao:2024split,Cao:2024universal} and single-insertion form factors \cite{Cao:2025ff,Zhang:2026FFsplit}. We further uncover two novel structures associated with operator insertions. A \emph{bulk zero} arises directly from the bulk integrations and has no counterpart in the corresponding boundary-only scattering amplitudes. It survives nonzero mutual bulk couplings: the common scale of the bulk punctures gives a scaleless integral, while their relative positions remain coupled. A \emph{decoupling split} distributes the operators between two nonempty subsets and produces two form-factor currents. This is a new possibility for multiple insertions: the usual $2$-split produces an amplitude current and a form-factor current, whereas a single insertion cannot supply operators to both factors. 

\section{String integrals for multiple insertions}
\label{sec:disk}
The general form factor for separate operator insertions is defined as the Fourier transform of a time-ordered product of operators,
\begin{align}
			F_{\mathcal{O}_1,\ldots,\mathcal{O}_m;n}={}&\int\prod_{\alpha=1}^{m}\dd^Dx_\alpha\,
			e^{-i\sum_\alpha q_{\alpha}\cdot x_\alpha}\notag\\
			\times&\langle p_1\cdots p_n|
			T\{\mathcal{O}_1(x_1)\cdots \mathcal{O}_m(x_m)\}|0\rangle ,
	\label{eq:multiff}
	\end{align}
where $p_i^2=0$ and $\sum_i p_i=\sum_\alpha q_\alpha$. We use connected, amputated color components, with couplings and the overall momentum delta function stripped. At tree level, we represent each operator in Eq.~\eqref{eq:multiff} by a distinct insertion vertex. Contributions in which several insertion labels meet at a single local vertex are treated separately as contact terms. Following Ref.~\cite{Cao:2025ff}, assign each $q_\alpha$ to a bulk puncture $z_{a_\alpha}\in\Hplus$ and its conjugate $z_{\bar a_\alpha}=\bar z_{a_\alpha}$. We label these coordinates by
\begin{equation}
 z_{a_\alpha}=z_{n+2\alpha-1},\qquad
 \bar z_{a_\alpha}=z_{n+2\alpha},\qquad \alpha=1,\ldots,m.
 \label{eq:bulklabels}
\end{equation}
The stringy general form factor is defined by
\begin{equation}
		\begin{split}
			\bF_{n;m}^{\{\mathcal{O}_\alpha\}}
			={}&\int_{\mathcal M_{n;m}^{\rm disk}}
			\frac{\prod_{i=1}^{n}\dd z_i}{\operatorname{vol}SL(2,\mathbb R)}
			\prod_{\alpha=1}^{m}
			\frac{\dd^2z_{a_\alpha}}
			{(z_{a_\alpha,\bar a_\alpha})^2}\\[-2pt]
			&\hspace{2.5em}\times
			\mathcal I_{n;m}^{\{\mathcal{O}_\alpha\}}\,\KN_{n;m} ,
		\end{split}
		\label{eq:diskdef}
	\end{equation}
Here $z_{r,s}=z_r-z_s$, and $\mathcal M_{n;m}^{\rm disk}$ has boundary order $z_1<\cdots<z_n$. We distinguish the dimensionless string invariants $s_{i,j}$ from the dimensionful invariants
\begin{equation}
 \hat{s}_{i,j}:=\frac{s_{i,j}}{\alpha'},\qquad i,j\in\{1,\ldots,n+2m\}.
 \label{eq:dimensionfulinvariants}
\end{equation}
The dimensionless invariants are analytically specialized to
\begin{equation}
		\begin{aligned}
			s_{i,j}&\longmapsto 2\alpha'p_i\cdot p_j,\qquad i,j\leq n,\\
			s_{i,a_\alpha}=s_{i,\bar a_\alpha}
			&\longmapsto-\alpha'p_i\cdot q_{\alpha},~~ s_{a_\alpha,\bar a_\alpha}\longmapsto\alpha'q_{\alpha}^2,\\
			s_{a_\alpha,a_\beta}=s_{\bar a_\alpha,\bar a_\beta}
			&=s_{a_\alpha,\bar a_\beta}\longmapsto\frac{\alpha'}2q_{\alpha}\cdot q_\beta,
			\quad\alpha\ne\beta .
		\end{aligned}
		\label{eq:generalimages}
	\end{equation}
All four invariants between distinct image pairs take the value in the last line. For an operator or auxiliary momentum $Q$, we also write $\hat{s}_{i,Q}=2p_i\cdot Q$ and $s_{i,Q}=\alpha'\hat{s}_{i,Q}$. In particular, $\hat{s}_{i,q_\alpha}=-2\hat{s}_{i,a_\alpha}=-2\hat{s}_{i,\bar a_\alpha}$. The Koba--Nielsen factor \cite{KobaNielsen:1969} is $\KN_{n;m}=\prod_{r<s}z_{r,s}^{s_{r,s}}$, on the branch inherited from the disk. Momentum conservation makes every row of the invariant matrix sum to zero, so $\KN_{n;m}$ has zero M\"obius weight. The off-shell assignment in Eq.~\eqref{eq:generalimages} defines an analytic family of integrals whose field-theory content we determine below; it is not an on-shell closed-string amplitude.

We consider $m$ identical local operators, $\mathcal{O}_1=\cdots=\mathcal{O}_m=\mathcal{O}$. In this construction, the external particle content determines the choice of inserted operator. For purely scalar external states, we take $\mathcal O=\mathcal L_{\phi^3}=\tfrac12\Tr(\partial_\mu\phi\,\partial^\mu\phi)+\tfrac{1}{3!}\Tr(\phi^3)$, with the coupling stripped. For purely gluonic external states, we take $\mathcal O=\mathcal O_{\rm YM}=\Tr(F^2)$, while for mixed scalar--gluon external states we take $\mathcal O=\mathcal L_{\rm YMS}$. The YMS Lagrangian and operator normalizations are specified in Appendix~\ref{app:wick-assembly}. The corresponding disk kernels are
\begin{align}
		\mathcal I_{n;m}^{\mathcal L_{\phi^3}}&=\PT(1,2,\ldots,n):=\frac{1}{z_{1,2}z_{2,3}\cdots z_{n,1}},
		\label{eq:kernelphi3}\\
		\mathcal I_{n;m}^{\mathcal O_{\rm YM}}&=\sum_{\mathcal M}\prod_{(i,j)\in\mathcal M}W_{i,j}\prod_{\ell\notin\mathcal M}V_\ell ,
		\label{eq:kernelYM}\\
		\mathcal I_{g,r;m}^{\mathcal L_{\rm YMS}}&=\prod_{k=1}^{t}\PT(\rho^{(k)})\sum_{\mathcal M\subset\{1,\ldots,g\}}
		\prod_{(i,j)\in\mathcal M}W_{i,j}\prod_{\ell\notin\mathcal M}V_\ell ,
		\label{eq:kernelYMS}
	\end{align}
	for purely scalar states, purely gluon states, and mixed states with $g$ gluons and $r$ scalars ($g+r=n$). The disjoint cyclic words $\rho^{(1)},\ldots,\rho^{(t)}$ partition the scalar labels into $t$ flavor traces. Here $\mathcal M$ runs over partial matchings of the gluon labels, and $\ell\notin\mathcal M$ denotes unmatched labels. The contractions use the assignment $\epsilon_i\cdot p_{a_\alpha}=\epsilon_i\cdot p_{\bar a_\alpha}=-\epsilon_i\cdot q_\alpha/2$ and are
\begin{equation}
\begin{aligned}
W_{i,j}=\frac{2\epsilon_i\cdot\epsilon_j}{\alpha' z_{i,j}^2},~~V_i=\sum_{\substack{j=1\\j\ne i}}^{n+2m}\frac{2\epsilon_i\cdot p_j}{z_{i,j}}
\end{aligned}
\label{eq:VWgeneral}
\end{equation}
Only $V_i$ depends on the bulk coordinates. The gluon splits and zeros below require additional polarization conditions on the Wick contractions.

\begin{figure}[t]
		\centering
		\begin{tikzpicture}[x=1cm,y=1cm,every node/.style={font=\scriptsize},boundary/.style={circle,fill=black,inner sep=1.7pt},pairA/.style={circle,fill=blue!70!black,inner sep=1.9pt},pairB/.style={circle,fill=orange!85!black,inner sep=1.9pt}]
			\path[use as bounding box] (0,-.65) rectangle (8.1,6.55);
			\node[anchor=west,font=\scriptsize\bfseries] at (.05,6.35) {$m=2,\ n=2:\quad (a_1,\bar a_1)=(3,4),\ (a_2,\bar a_2)=(5,6)$};
			\node[anchor=west] at (.05,6.05) {(a) $w=(1,2,3,4,5,6)$};
			\draw[black!45] (.25,5.35)--(7.35,5.35);
			\node[boundary] at (0.6,5.35) {};
			\node[below=3pt] at (0.6,5.35) {$1$};
			\node[boundary] at (1.88,5.35) {};
			\node[below=3pt] at (1.88,5.35) {$2$};
			\node[pairA] at (3.16,5.35) {};
			\node[below=3pt] at (3.16,5.35) {$3$};
			\node[pairA] at (4.4399999999999995,5.35) {};
			\node[below=3pt] at (4.4399999999999995,5.35) {$4$};
			\node[pairB] at (5.72,5.35) {};
			\node[below=3pt] at (5.72,5.35) {$5$};
			\node[pairB] at (7.0,5.35) {};
			\node[below=3pt] at (7.0,5.35) {$6$};
			\node[anchor=west] at (.6,4.71) {$\Phi_{2,2}(w)=\textcolor{blue!70!black}{s_{3,4}}+\textcolor{orange!85!black}{s_{5,6}}$};
			\node[anchor=west] at (.05,3.8499999999999996) {(b) $w=(1,3,2,4,5,6)$};
			\fill[blue!7] (1.88,2.91) rectangle (4.4399999999999995,3.19);
			\draw[black!45] (.25,3.15)--(7.35,3.15);
			\node[boundary] at (0.6,3.15) {};
			\node[below=3pt] at (0.6,3.15) {$1$};
			\node[pairA] at (1.88,3.15) {};
			\node[below=3pt] at (1.88,3.15) {$3$};
			\node[boundary] at (3.16,3.15) {};
			\node[below=3pt] at (3.16,3.15) {$2$};
			\node[pairA] at (4.4399999999999995,3.15) {};
			\node[below=3pt] at (4.4399999999999995,3.15) {$4$};
			\node[pairB] at (5.72,3.15) {};
			\node[below=3pt] at (5.72,3.15) {$5$};
			\node[pairB] at (7.0,3.15) {};
			\node[below=3pt] at (7.0,3.15) {$6$};
			\node[anchor=west] at (.6,2.51) {$\Phi_{2,2}(w)=\textcolor{blue!70!black}{s_{3,4}}+\textcolor{orange!85!black}{s_{5,6}}+s_{2,3}$};
			\node[anchor=west,text=black!65] at (.6,2.13) {$3<_w2<_w4:\quad +s_{2,3}$};
			\node[anchor=west] at (.05,1.65) {(c) $w=(1,3,2,5,4,6)$};
			\fill[blue!7] (1.88,0.71) rectangle (5.72,0.99);
			\draw[black!45] (.25,0.95)--(7.35,0.95);
			\node[boundary] at (0.6,0.95) {};
			\node[below=3pt] at (0.6,0.95) {$1$};
			\node[pairA] at (1.88,0.95) {};
			\node[below=3pt] at (1.88,0.95) {$3$};
			\node[boundary] at (3.16,0.95) {};
			\node[below=3pt] at (3.16,0.95) {$2$};
			\node[pairB] at (4.4399999999999995,0.95) {};
			\node[below=3pt] at (4.4399999999999995,0.95) {$5$};
			\node[pairA] at (5.72,0.95) {};
			\node[below=3pt] at (5.72,0.95) {$4$};
			\node[pairB] at (7.0,0.95) {};
			\node[below=3pt] at (7.0,0.95) {$6$};
			\node[anchor=west] at (.6,0.30999999999999994) {$\Phi_{2,2}(w)=\textcolor{blue!70!black}{s_{3,4}}+\textcolor{orange!85!black}{s_{5,6}}+s_{2,3}+\textcolor{violet}{s_{3,6}}$};
			\node[anchor=west,text=violet] at (.6,-0.07000000000000006) {$3<_w5<_w4<_w6:\quad +s_{3,6}$};
		\end{tikzpicture}
		\caption{Continuation phases for boundary labels $1,2$ and bulk image pairs $(3,4)$ and $(5,6)$. Black points denote boundary labels; blue and orange points denote the image pairs. Shading marks the interval between $3$ and $4$. Each row displays the full phase, with chamber weight $e^{i\pi\Phi_{2,2}(w)}$. Here $r<_ws$ means that label $r$ precedes label $s$ in the ordering $w$.}
		\label{fig:phase-unfolding}
	\end{figure}
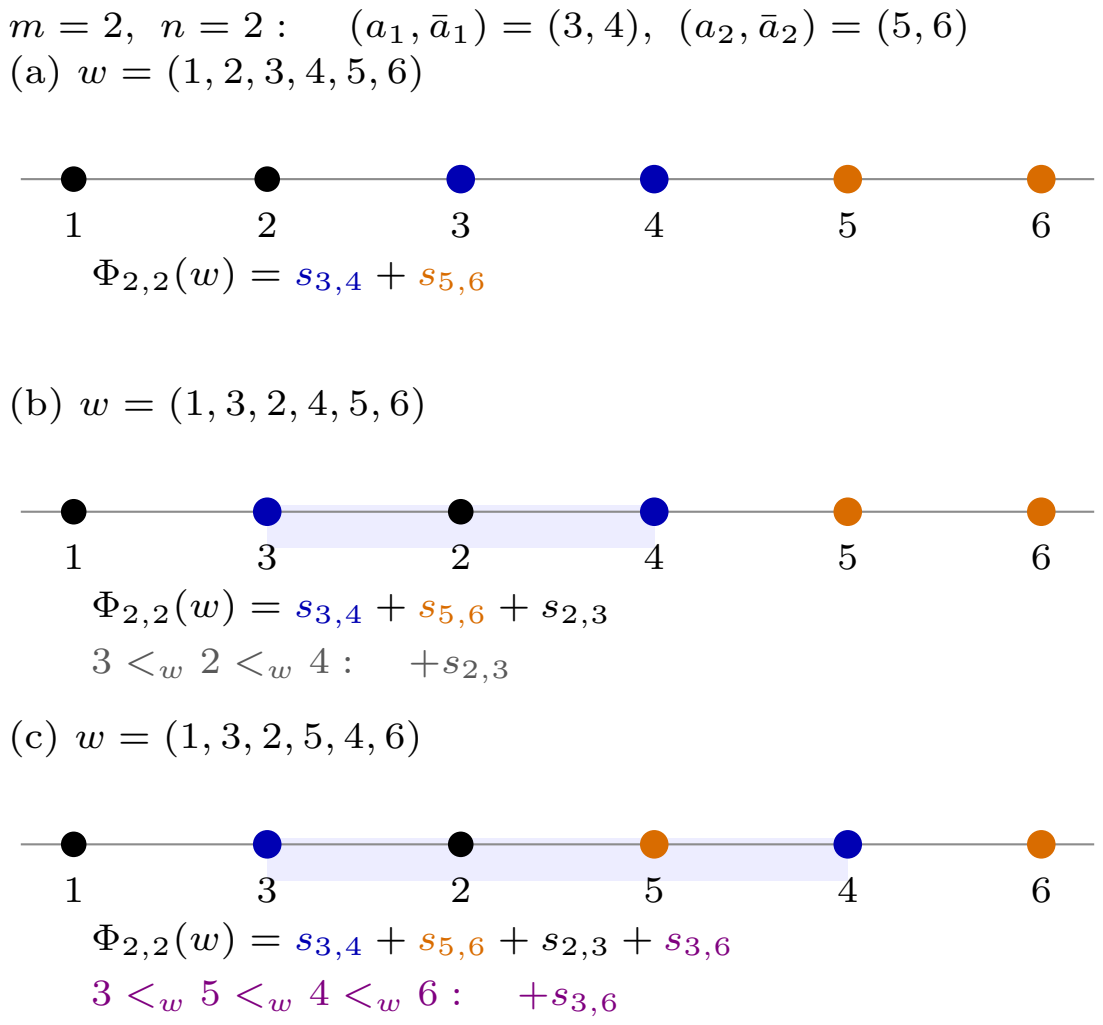

\section{The field-theory limit}
\label{sec:field-theory}
Continuing each conjugate bulk pair to independent real positions rewrites the disk integrals as sums of ordered boundary integrals \cite{Stieberger:2009hq}. This representation makes the monodromy phases entering the field-theory limit explicit. For an ordering $w=(w_1,\ldots,w_{n+2m})$, let $\pos_w(r)$ denote the position of label $r$ in $w$, so that $\pos_w(w_\ell)=\ell$. With label $1$ fixed, the allowed chambers preserve the relative order of the boundary labels $(2,\ldots,n)$: $\cP_{n,m}=\{w=(1,\sigma):\pos_w(2)<\cdots<\pos_w(n)\}$, with $|\cP_{n,m}|=(n+2m-1)!/(n-1)!$.
Here $\sigma$ permutes $\{2,\ldots,n+2m\}$. For a complete scalar, gluon, or mixed kernel $I$, define
\begin{equation}
A_{n+2m}^{(I)}(w)=\int_{D(w)}\dd\mu_{n+2m}
\prod_{i<j}|x_i-x_j|^{s_{i,j}}\,\mathcal R^{(I)}(x),
\label{eq:orderedA}
\end{equation}
where $D(w)=\{x_{w_1}<\cdots<x_{w_{n+2m}}\}$ and $\mathcal R^{(I)}$ includes the image-pair factors and all Wick contractions. For $x<0$, $(x\pm i0)^s=|x|^s e^{\pm i\pi s}$, so branch crossings give a constant phase in each chamber. Integer powers retain their signs in the rational kernel. Combining the chamber phases with the contour Jacobian expresses the stringy general form factor in terms open-string
amplitudes,
\begin{equation}
			\bF_{n;m}^{(I)}
			=\left(\frac{i}{2}\right)^m
			\sum_{w\in\cP_{n,m}}
			e^{i\pi\Phi_{n,m}(w)}A_{n+2m}^{(I)}(w) .
		\label{eq:master}
	\end{equation}
\begin{samepage}
The phase $\Phi_{n,m}$ records self-image, boundary--bulk, and bulk--bulk crossings. On the branch of Ref.~\cite{Stieberger:2009hq}, Eq.~(2.17), with $x_1=-\infty$, define $\theta_w(r>s)=\mathbf1_{\pos_w(r)>\pos_w(s)}$ and $C_w(r,s;t,u)=\theta_w(r>s)\oplus\theta_w(t>u)$, where $u\oplus v=u+v-2uv$. Then\par
\end{samepage}
\begin{equation}
\begin{aligned}
\Phi_{n,m}(w)=&\sum_{\alpha=1}^m s_{a_\alpha,\bar a_\alpha}\theta_w(\bar a_\alpha>a_\alpha)\\
&+\sum_{i=2}^n\sum_{\alpha=1}^m s_{i,a_\alpha}C_w(i,a_\alpha;i,\bar a_\alpha)\\
&+\sum_{\alpha<\beta}s_{a_\alpha,a_\beta}C_w(a_\alpha,a_\beta;\bar a_\alpha,\bar a_\beta)\\
&+\sum_{\alpha<\beta}s_{a_\alpha,\bar a_\beta}C_w(a_\alpha,\bar a_\beta;\bar a_\alpha,a_\beta).
\end{aligned}\label{eq:generalphase}
\end{equation}

Figure~\ref{fig:phase-unfolding} shows how the phase is read from an ordering for $n=m=2$. All three chambers retain the self-image terms $s_{3,4}+s_{5,6}$. Placing boundary label $2$ between $3$ and $4$ adds the boundary--bulk term $s_{2,3}$ in (b); interleaving the two image pairs adds the bulk--bulk term $s_{3,6}$ in (c). Interactions between distinct insertions thus enter the continuation phase directly.

The continuation phases approach unity as $\alpha'\to0$, but they multiply ordered integrals that can develop poles. Their expansion must therefore be retained when extracting the field-theory coefficient. Choose $r_I$ to be the largest pole order among the ordered integrals of the complete kernel $I$, so that
\begin{align}\label{eq:normalizedexpansion}
			\widehat A_{n+2m}^{(I)}(w)&:=(\alpha')^{r_I}A_{n+2m}^{(I)}(w)
			=A_{n+2m}^{(I,0)}(w)+O(\alpha'),\notag\\
			\Phi_{n,m}(w)&=\alpha'\widehat\Phi_{n,m}(w).
		\end{align}
\begin{samepage}
For $n\geq2$, monodromy cancels all terms below order $(\alpha')^m$ in the normalized sum. At that order only the $m$th phase moment remains:\par
\end{samepage}
\begin{equation}
			\left.\bF_{n;m}^{(I)}\right|_{\rm FT}
			=\frac{(-1)^m\pi^m}{2^m m!}
			\sum_{w\in\cP_{n,m}}
			\widehat\Phi_{n,m}(w)^m A_{n+2m}^{(I,0)}(w) .
		\label{eq:mthmoment}
	\end{equation}
Here $\left.\bF^{(I)}\right|_{\rm FT}:=\lim_{\alpha'\to0}(\alpha')^{r_I-m}\bF^{(I)}$. The phase and Laurent series must be expanded together, as shown in Appendix~\ref{app:wick-assembly}. The subleading Laurent coefficients cancel from this order of the normalized sum. The field-theory coefficient thus requires only leading multitrace YMS amplitudes \cite{He:2019drm}, in the worldsheet normalization of Eqs.~\eqref{eq:orderedA} and \eqref{eq:VWgeneral}. We denote them by $A^{\rm YMS}_{n+2m}(\tau_1|\cdots|\tau_r;G\Vert w)$, with scalar traces $\tau_j$, gluons $G$, and gauge ordering $w$. With $\tau_0=(1,\ldots,n)$ and $\mathcal T_m=(n+1,n+2)|\cdots|(n+2m-1,n+2m)$,
\begin{equation}
\begin{aligned}
A_{n+2m}^{(\mathcal L_{\phi^3},0)}(w)&=A_{n+2m}^{\rm YMS}(\tau_0|\mathcal T_m;\varnothing\Vert w),\\
A_{n+2m}^{(\mathcal O_{\rm YM},0)}(w)&=A_{n+2m}^{\rm YMS}(\mathcal T_m;\{1,\ldots,n\}\Vert w).
\end{aligned}
\label{eq:YMSleading}
\end{equation}
Each image pair supplies a two-particle scalar trace; the boundary labels form one further trace or remain gluons. The field-theory examples below use the cyclic amplitude convention specified in Appendix~\ref{app:wick-assembly}. There we give the pole counting, including the explicit $1/\alpha'$ factors in $W_{ij}$, and the conventions used with \texttt{Amp.m} \cite{HeHouTianZhang:2021}.

\section{Separated insertions and contact vertices}
\label{sec:contacts}
The phase-moment formula~\eqref{eq:mthmoment} gives the disk coefficient in terms of auxiliary amplitudes. Matching this coefficient to tree form factors reveals both propagation between distinct insertion vertices and local contacts in the examples below. These contributions can be organized by grouping the insertion labels:
\begin{equation}
			\left.\bF_{n;m}\right|_{\rm FT}
			=\sum_{\pi\in\Part([m])}C_\pi\,
			F_{\{\mathcal{O}^{(B)}(Q_B)\}_{B\in\pi};n}.
		\label{eq:partitionformula}
	\end{equation}
Here $\pi$ partitions $[m]=\{1,\ldots,m\}$ into blocks $B$, with $Q_B=\sum_{\alpha\in B}q_\alpha$. A singleton gives a separate insertion; a larger block gives a local operator carrying the combined momentum $Q_B$. A contact joins several insertions at one vertex; propagators may still connect that vertex to the external states. We establish this decomposition for $(n,m)=(2,2),(3,2),(2,3)$ with scalars and $(2,2)$ with gluons. In general, a merged block may require a sum of local operators rather than the original $\mathcal O$ alone; Eq.~\eqref{eq:partitionformula} is therefore not yet an arbitrary-multiplicity result.

The partition structure can be illustrated by a formal source expansion inspired by D-brane effective actions \cite{GarousiMyers:1999,Myers:1999}. Let $J$ be an external source and $\varphi$ the dynamical fields. Keeping terms without spacetime derivatives of $J$, we write schematically
\begin{equation}
		S_{\rm src}[\varphi;J]=S_0[\varphi]
		+\sum_{r\ge1}\frac1{r!}\int\dd^Dx\,J(x)^r \mathcal{O}^{(r)}(x).
		\label{eq:sourceaction}
	\end{equation}
Here $S_0$ is the action at $J=0$. In the resulting tree diagrams, source derivatives evaluated at $J=0$ assign insertion labels to vertices. Derivatives acting at distinct vertices give separate insertions; several derivatives acting at one vertex give a contact. Each block thus specifies the labels that meet at a vertex. This picture provides intuition for the partition structure; it does not establish a term-by-term correspondence with a specific D-brane action. In the examples below, the operator content and coefficients are determined by matching the disk coefficient to tree form factors. For two insertions,
\begin{equation}
			\left.\bF_{n;2}\right|_{\rm FT}
			=C_{1|2}F_{\mathcal{O},\mathcal{O};n}(q_a,q_b)
			+C_{12}F_{\mathcal{O}^{(2)};n}(q_a+q_b).
		\label{eq:fieldlimit}
	\end{equation}

We evaluate the disk side using Eq.~\eqref{eq:mthmoment} and the operator form factors using tree diagrams. For two particles and two insertions, $q_2=p_1+p_2-q_1$. We use the dimensionful invariants $\hat{s}_{i,q}=2p_i\cdot q$ and $\hat{s}_{1,2}=2p_1\cdot p_2$. Double lines denote operator insertions with incoming momentum $q_\alpha$ or $Q=q_1+q_2$; external momenta are outgoing. Straight and wavy lines denote scalars and gluons. For two scalars,
\begin{align}
F_{\mathcal O,\mathcal O;2}
&=\ffexchange{ffscalar}{1}{2}+\ffexchange{ffscalar}{2}{1}\notag\\
&=\frac{\hat{s}_{1,q_1}(\hat{s}_{1,2}-\hat{s}_{2,q_1})}{4(q_1^2-\hat{s}_{1,q_1})}
+\frac{(\hat{s}_{1,2}-\hat{s}_{1,q_1})\hat{s}_{2,q_1}}{4(q_1^2-\hat{s}_{2,q_1})},\notag\\
F_{\mathcal O;2}(Q)&=\ffcontact{ffscalar}=\frac{\hat{s}_{1,2}}2.
\label{eq:m2scalarcontact}
\end{align}
	\begin{equation}
		\left.\bF_{2;2}^{\mathcal L_{\phi^3}}\right|_{\rm FT}
			=\pi^2\left[
			F_{\mathcal{O},\mathcal{O};2}(q_1,q_2)+F_{\mathcal{O};2}(q_1+q_2)\right].
		\label{eq:m2scalardecomposition}
	\end{equation}
	The poles $(q_1-p_i)^2=0$ describe particles propagating between the two insertion vertices. In this two-particle example, the merged contribution is polynomial and carries only the total operator momentum $Q$. The disk coefficient contains both contributions.
	
We test the same decomposition for gluons, keeping their polarization dependence explicit. Write $\Tr(\alpha_1,\ldots,\alpha_r)=\Tr(f_{\alpha_1}\cdots f_{\alpha_r})$ and $\operatorname{tr}(\alpha_1,\ldots,\alpha_r)=q_1\cdot f_{\alpha_1}\cdots f_{\alpha_r}\cdot q_2$, with $f_i^{\mu\nu}=p_i^\mu\epsilon_i^\nu-p_i^\nu\epsilon_i^\mu$. The diagrams give
\begin{align}
F_{\mathcal O,\mathcal O;2}
&=\ffexchange{ffgluon}{1}{2}+\ffexchange{ffgluon}{2}{1}\notag\\
&=-\frac{\operatorname{tr}(1,2)}{\hat{s}_{1,q_1}-q_1^2}
-\frac{\operatorname{tr}(2,1)}{\hat{s}_{2,q_1}-q_1^2},\label{eq:m2YMsep}\\
F_{\mathcal O;2}(Q)&=\ffcontact{ffgluon}=\frac{\Tr(1,2)}2.
\label{eq:m2YMcontact}
\end{align}
	\begin{equation}
		\left.\bF_{2;2}^{\mathcal O_{\rm YM}}\right|_{\rm FT}
			=2\pi^2\left[
			F_{\mathcal{O},\mathcal{O};2}(q_1,q_2)+F_{\mathcal{O};2}(q_1+q_2)\right].
		\label{eq:m2YMdecomposition}
	\end{equation}
The factor of two accounts for the normalization of the gluon disk kernel relative to the tree vertices. Appendix~\ref{app:wick-assembly} checks this factor and gives the scalar examples with $(n,m)=(3,2)$ and $(2,3)$. Three insertions allow both partial and complete mergers: two labels can meet at one vertex while the third remains separate, or all three can meet together.

\section{Hidden structures}
\label{sec:splits-zeros}
At finite $\alpha'$, the stringy general form factors exhibit several hidden structures. Their $2$-splits and hidden zeros extend the corresponding structures of scattering amplitudes. We also find two novel structures of stringy general form factors. The first is a decoupling split into two form-factor currents, a feature unique to multiple operator insertions. The second is the emergence of bulk zeros associated with operator punctures, which have no counterpart in boundary-only scattering amplitudes.

\subsection{2-split}
\label{sec:two-split}
We first consider a split in which one current contains all operator insertions, extending the single-insertion split of Ref.~\cite{Cao:2025ff}. Choose boundary labels $i,j,k$. Let $A$ contain the labels on the arc from $i$ to $j$ that excludes $k$, and let $B$ contain the remaining non-shared labels. Assign every bulk puncture to the $B$ side and impose
\begin{equation}
		s_{a,b}=s_{a,q_{\alpha}}=0,\qquad
		a\in A,\quad
		b\in B, \quad q_{\alpha}\in\mathsf Q_{m}.
		\label{eq:sameside}
	\end{equation}
Here $\mathsf Q_m=\{q_\alpha:\alpha\in [m]\}$; each condition involving $q_\alpha$ is imposed on both image labels through Eq.~\eqref{eq:generalimages}. On this locus, denoted $\mathcal K_{\rm same}$,
\begin{equation}
			\bF_{n;m}^{\mathcal L_{\phi^3}}\big|_{\mathcal K_{\rm same}}
			=\bJ^{\phi^3}_{|A|+3}(i,A,j;\kappa_A)
			\times\bF_{|B|+3;m}^{\mathcal L_{\phi^3}}
			(j,B,i;\kappa_B).
		\label{eq:samesidesplit}
	\end{equation}
The factors on the right of Eq.~\eqref{eq:samesidesplit} are off-shell currents defined by the inherited gauge-fixed integrals. Their auxiliary legs replace $k$ at infinity; conservation fixes $\kappa_A,\kappa_B$, which need not be null. We use $\bJ$ for an amplitude current with no operator insertions and $\bF$ for a form-factor current containing insertions. The disk current $\bJ$ retains string corrections; its normalized leading coefficient is the cubic-scalar current. All boundary legs are scalar in Eq.~\eqref{eq:samesidesplit}, so no particle-type superscripts are needed.

With $z_i=0,z_j=1,z_k=\infty$, every term joining the two groups drops out of $\log\KN$. The Parke--Taylor chains, measure, and domain then separate \cite{Cao:2024split,Cao:2024universal}. This proves the identity at finite $\alpha'$, without taking a propagator residue.

For $\mathcal O_{\rm YM}$, the momentum conditions alone leave cross contractions in the Wick kernel. These vanish if, in addition,
\begin{equation}
p_a\cdot\epsilon_{b'}=\epsilon_a\cdot\epsilon_{b'}=\epsilon_a\cdot p_b=\epsilon_a\cdot q_\alpha=0,
\label{eq:splitpolarizations}
\end{equation}
where $a\in A$, $b\in B$, $b'\in B\cup\{i,j,k\}$, and $\alpha\in[m]$. Denote the combined locus by $\mathcal K_{\rm same}^{\rm YM}$. Then
\begin{equation}
\begin{split}
\left.\bF_{n;m}^{\mathcal O_{\rm YM}}\right|_{\mathcal K_{\rm same}^{\rm YM}}
={}&\bJ^{\rm mixed}_{|A|+3}(i^\phi,A,j^\phi;\kappa_A^\phi)\\
&\times\bF^{\mathcal O_{\rm YM}}_{|B|+3;m}(j,B,i;\kappa_B)_\mu\epsilon_k^\mu.
\end{split}
\label{eq:samesidesplitYM}
\end{equation}
Here $\bJ^{\rm mixed}$ has gluons in $A$ and three scalar legs $i,j,\kappa_A$, marked by the superscript $\phi$ to distinguish them from gluons. The right current carries the polarizations of $B\cup\{i,j\}$ and an open Lorentz index for the auxiliary leg, contracted with $\epsilon_k$, as in the single-insertion split \cite{Cao:2025ff}.

\subsection{Hidden zeros}
\label{sec:hidden-zeros}
The $2$-split also yields hidden zeros when one current vanishes and the other remains regular. We apply this mechanism to Eq.~\eqref{eq:samesidesplit}, whose left current contains no operator insertions. Let $Q=\sum_\alpha q_\alpha$ and $p_B=\sum_{b\in B}p_b$. Then
\begin{equation}
		\kappa_A=p_k+p_B-Q=-p_i-p_j-\sum_{a\in A}p_a.
		\label{eq:inheritedkappa}
	\end{equation}
On $\mathcal K_{\rm same}$, $2\alpha'p_a\cdot\kappa_A=s_{a,k}$. For nonempty $A$, the additional conditions
\begin{equation}
		s_{a,k}=0\qquad\text{for all }a\in A
		\label{eq:inheritedzero}
	\end{equation}
therefore place the scalar current on its known hidden zero \cite{ArkaniHamed:2023zero,Cao:2024universal}. The product vanishes wherever the second factor is regular. For the polarized split, imposing also $\epsilon_a\cdot p_k=0$ puts the left mixed current in Eq.~\eqref{eq:samesidesplitYM} on its hidden zero \cite{Cao:2025ff}. Hence, away from poles,
\begin{equation}
\left.\bF_{n;m}^{\mathcal O_{\rm YM}}\right|_{\substack{\mathcal K_{\rm same}^{\rm YM}\\ s_{a,k}=\epsilon_a\cdot p_k=0\ (a\in A)}}=0.
\label{eq:hiddenzeroYM}
\end{equation}

\subsection{Decoupling split}
\label{sec:decoupling-split}
Multiple insertions admit a new split in which both factors contain operators. This \emph{decoupling split} produces two form-factor currents, in contrast to the amplitude current and form-factor current in the $2$-split above. It requires at least two insertions and hence has no single-insertion counterpart. Write $[m]=M_A\sqcup M_B$ with $M_A,M_B\ne\varnothing$, keeping each complete image pair on its respective side. Define $\mathsf Q_M=\{q_\alpha:\alpha\in M\}$ and $Q_M=\sum_{\alpha\in M}q_\alpha$. The required locus $\mathcal K_{\rm dec}$ is
\begin{equation}
		\begin{split}
			s_{\tilde{a},\tilde{b}}=0,\quad
			\tilde{a}\in A\cup\mathsf Q_{M_A},~\tilde{b}\in B\cup\mathsf Q_{M_B}.
		\end{split}
		\label{eq:oppositeside}
	\end{equation}
This sets every coupling between the two groups to zero, including couplings between their bulk punctures. The factorization argument of Sec.~\ref{sec:two-split} then gives
\begin{equation}
		\begin{split}
			\bF_{n;m}^{\mathcal L_{\phi^3}}\big|_{\mathcal K_{\rm dec}}
			={}&\bF_{|A|+3;|M_A|}^{\mathcal L_{\phi^3}}
			(i,A,j;\kappa_A;\mathsf Q_{M_A})\\[-2pt]
			&\times\bF_{|B|+3;|M_B|}^{\mathcal L_{\phi^3}}
			(j,B,i;\kappa_B;\mathsf Q_{M_B}).
		\end{split}
		\label{eq:sepsplit}
	\end{equation}
Figure~\ref{fig:decoupling-split} illustrates this split in a disk representation. Both factors are off-shell stringy form-factor currents defined by the inherited gauge-fixed disk integrals. The first subscript counts the boundary legs, including the auxiliary leg; the second counts the operator insertions. The ordered arguments specify the boundary legs, and the final set lists the operator momenta assigned to that current. The auxiliary legs replace $k$ at infinity, with momenta fixed by
\begin{equation}
\begin{aligned}
\kappa_A&=Q_{M_A}-p_i-p_j-\sum_{a\in A}p_a,\\
\kappa_B&=Q_{M_B}-p_i-p_j-\sum_{b\in B}p_b.
\end{aligned}
\label{eq:decouplingkappa}
\end{equation}
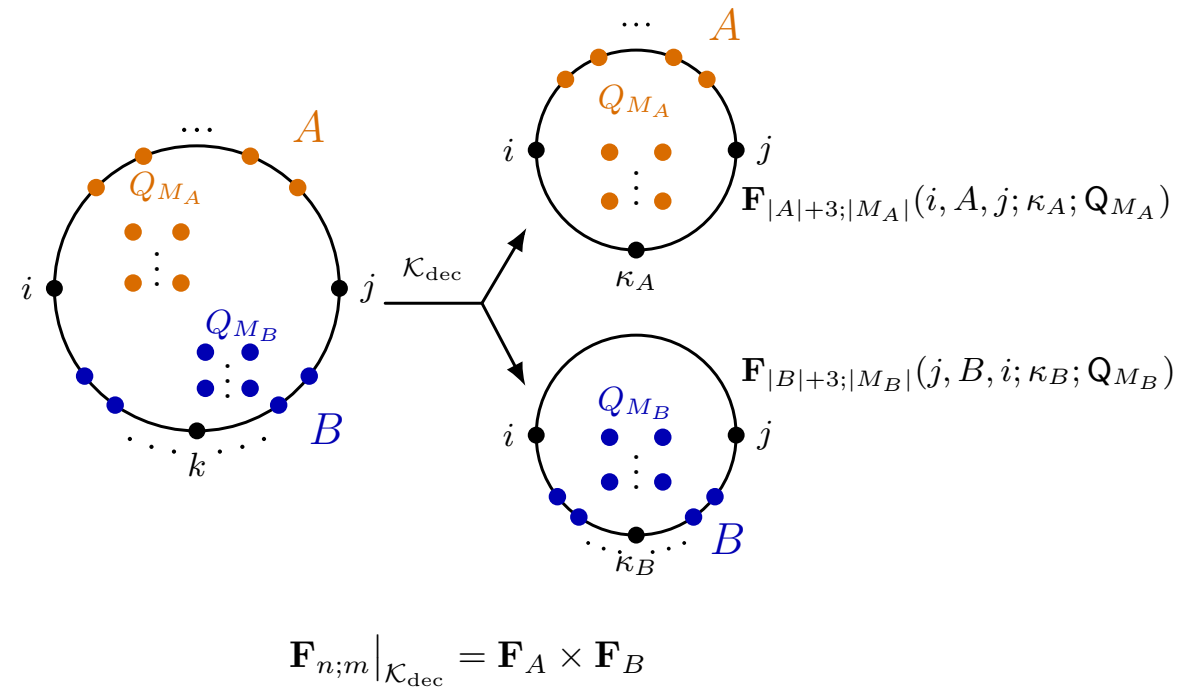
\begin{figure}[t]
  \centering
  \resizebox{0.91\columnwidth}{!}{%
    \begin{tikzpicture}[
     x=1cm,y=1cm,line cap=round,
     every node/.style={font=\scriptsize},
     rim/.style={line width=.8pt},
     bd/.style={circle,fill=black,inner sep=1.6pt},
     Adot/.style={circle,fill=orange!85!black,inner sep=1.65pt},
     Bdot/.style={circle,fill=blue!70!black,inner sep=1.65pt},
     Atext/.style={text=orange!85!black},
     Btext/.style={text=blue!70!black},
     split/.style={line width=.8pt,-{Latex[length=2.2mm,width=1.7mm]}},
     edots/.style={font=\normalsize}
    ]
    \coordinate (L) at (0,0);
    \def\rL{1.34}
    \draw[rim] (L) circle (\rL);
    \node[bd,label={[font=\small]left:$i$}] at ($(L)+(180:\rL)$) {};
    \node[bd,label={[font=\small]right:$j$}] at ($(L)+(0:\rL)$) {};
    \node[bd,label={[font=\small]below:$k$}] at ($(L)+(270:\rL)$) {};
    \foreach \a in {135,112,68,45}{\node[Adot] at ($(L)+(\a:\rL)$) {};}
    \foreach \a in {218,235,305,322}{\node[Bdot] at ($(L)+(\a:\rL)$) {};}
    \foreach \dx in {-0.12,0,0.12}{\fill ($(0,1.49)+(\dx,0)$) circle (.020cm);}    
    \node[Atext,font=\large] at (1.05,1.52) {$A$};
    \node[Btext,font=\large] at (1.22,-1.32) {$B$};
    \foreach \ang in {246,254,262,278,286,294}{\fill ($(L)+(\ang:1.55)$) circle (.020cm);}    
    \node[Atext,font=\small] at (-.28,.94) {$Q_{M_A}$};
    \foreach \x in {-.60,-.15}{\foreach \y in {.53,.05}{\node[Adot] at (\x,\y) {};}}
    \node[edots] at (-.375,.29) {$\vdots$};
    \node[Btext,font=\small] at (.44,-.35) {$Q_{M_B}$};
    \foreach \x in {.08,.50}{\foreach \y in {-.60,-.94}{\node[Bdot] at (\x,\y) {};}}
    \node[edots] at (.29,-.77) {$\vdots$};
    
    \coordinate (J) at (2.68,-.14);
    \draw[rim] (1.77,-.14) -- (J);
    \node[above=3pt] at (2.22,-.14) {$\mathcal K_{\mathrm{dec}}$};
    \draw[split] (J)--(3.10,.57);
    \draw[split] (J)--(3.09,-.92);
    
    \coordinate (UA) at (4.13,1.30);
    \def\rR{0.94}
    \draw[rim] (UA) circle (\rR);
    \node[bd,label={[font=\small]left:$i$}] at ($(UA)+(180:\rR)$) {};
    \node[bd,label={[font=\small]right:$j$}] at ($(UA)+(0:\rR)$) {};
    \node[bd,label={[font=\small]below:$\kappa_A$}] at ($(UA)+(270:\rR)$) {};
    \foreach \a in {135,112,68,45}{\node[Adot] at ($(UA)+(\a:\rR)$) {};}
    \foreach \dx in {-0.11,0,0.11}{\fill ($(4.13,2.48)+(\dx,0)$) circle (.018cm);}    
    \node[Atext,font=\large] at (4.96,2.48) {$A$};
    \node[Atext,font=\small] at (4.13,1.76) {$Q_{M_A}$};
    \foreach \x in {3.88,4.38}{\foreach \y in {1.28,.82}{\node[Adot] at (\x,\y) {};}}
    \node[edots] at (4.13,1.05) {$\vdots$};
    \node[anchor=west,font=\small] at (5.,0.8) {$\mathbf F_{|A|+3;|M_A|}(i,A,j;\kappa_A;\mathsf Q_{M_A})$};
    
    \coordinate (UB) at (4.13,-1.38);
    \draw[rim] (UB) circle (\rR);
    \node[bd,label={[font=\small]left:$i$}] at ($(UB)+(180:\rR)$) {};
    \node[bd,label={[font=\small]right:$j$}] at ($(UB)+(0:\rR)$) {};
    \node[bd,label={[font=\small]below:$\kappa_B$}] at ($(UB)+(270:\rR)$) {};
    \foreach \a in {218,235,305,322}{\node[Bdot] at ($(UB)+(\a:\rR)$) {};}
    \foreach \ang in {246,254,262,278,286,294}{\fill ($(UB)+(\ang:1.11)$) circle (.018cm);}    
    \node[Btext,font=\large] at (4.99,-2.35) {$B$};
    \node[Btext,font=\small] at (4.13,-1.08) {$Q_{M_B}$};
    \foreach \x in {3.88,4.38}{\foreach \y in {-1.40,-1.82}{\node[Bdot] at (\x,\y) {};}}
    \node[edots] at (4.13,-1.62) {$\vdots$};
    \node[anchor=west,font=\small] at (5.,-0.8) {$\mathbf F_{|B|+3;|M_B|}(j,B,i;\kappa_B;\mathsf Q_{M_B})$};
    
    \node[font=\normalsize] at (2.55,-3.5)
     {$\displaystyle \mathbf F_{n;m}\big|_{\mathcal K_{\mathrm{dec}}}=\mathbf F_A\times\mathbf F_B$};
    \end{tikzpicture}
  }
  \caption{Disk representation of the decoupling split on $\mathcal K_{\rm dec}$. Orange points represent boundary legs in $A$ and insertions in $M_A$; blue points represent boundary legs in $B$ and insertions in $M_B$. The bulk labels $Q_{M_A}$ and $Q_{M_B}$ denote total operator momenta, while $\mathsf Q_{M_A}$ and $\mathsf Q_{M_B}$ in the current arguments denote the corresponding sets. The two factors are form-factor currents with auxiliary legs $\kappa_A$ and $\kappa_B$, as in Eq.~\eqref{eq:sepsplit}.}
  \label{fig:decoupling-split}
\end{figure}
Neither auxiliary momentum is required to be null. The scalar currents in Eq.~\eqref{eq:sepsplit} contain the insertions of $\mathcal L_{\phi^3}$ in their respective sets. If $M_A=\varnothing$, the left factor contains no insertions and reduces to the amplitude current $\bJ^{\phi^3}$, recovering Eq.~\eqref{eq:samesidesplit}.

For $\mathcal O_{\rm YM}$, supplement $\mathcal K_{\rm dec}$ with
\begin{equation}
\epsilon_a\cdot p_{\tilde b}=\epsilon_{b'}\cdot p_{\tilde a}=\epsilon_a\cdot\epsilon_{b'}=0,
\label{eq:decpolarizations}
\end{equation}
where $a\in A$, $b'\in B\cup\{i,j,k\}$, $\tilde a\in A\cup\mathsf Q_{M_A}$, and $\tilde b\in B\cup\mathsf Q_{M_B}$. On this combined locus $\mathcal K_{\rm dec}^{\rm YM}$,
\begin{equation}
\begin{split}
\left.\bF_{n;m}^{\mathcal O_{\rm YM}}\right|_{\mathcal K_{\rm dec}^{\rm YM}}
={}&\bF^{\rm mixed}_{|A|+3;|M_A|}
(i^\phi,A,j^\phi;\kappa_A^\phi;\mathsf Q_{M_A})\\
&\times\bF^{\mathcal O_{\rm YM}}_{|B|+3;|M_B|}
(j,B,i;\kappa_B;\mathsf Q_{M_B})_\mu\epsilon_k^\mu.
\end{split}
\label{eq:sepsplitYM}
\end{equation}
The left factor $\bF^{\rm mixed}$ is the inherited mixed scalar--gluon form-factor current, with gluons in $A$, scalar legs $i,j,\kappa_A$, and the bulk insertions in $M_A$. The right form-factor current contains the insertions in $M_B$, carries the polarizations of $B\cup\{i,j\}$, and has an open Lorentz index for $\kappa_B$, contracted with $\epsilon_k^\mu$. Both currents use the auxiliary momenta in Eq.~\eqref{eq:decouplingkappa}. Appendix~\ref{app:splitlemma-full} derives both polarized identities.

For mixed kernels, each flavor-trace Parke--Taylor factor must respect the split: no adjacent pair, including the closing pair, may join non-shared scalars on opposite sides. Appendix~\ref{app:splitlemma-full} gives a four-point counterexample with a crossing trace.

\subsection{Bulk zeros}
\label{sec:bulk-zeros}
We now identify a zero intrinsic to the operator insertions. Unlike the hidden zeros in Sec.~\ref{sec:hidden-zeros}, this zero follows directly from the bulk integration, rather than from a vanishing amplitude current. We call it a \emph{bulk zero}; its mechanism has no counterpart in the corresponding boundary-only scattering amplitudes.

For $n\geq3$ and $m\geq1$, choose distinct boundary labels $i,j$ and require every operator momentum to couple only to these two boundary momenta:
\begin{equation}
 p_r\cdot q_\alpha=0,\qquad
 r\notin\{i,j\},\quad \alpha\in[m].
 \label{eq:bulkzerolocus}
\end{equation}
On this locus, denoted $\mathcal K_{\rm bulk}^{(i,j)}$, the scalar integral vanishes away from poles:
\begin{equation}
 \left.\bF_{n;m}^{\mathcal L_{\phi^3}}
 \right|_{\mathcal K_{\rm bulk}^{(i,j)}}=0.
 \label{eq:bulkzeroscalar}
\end{equation}

To expose the mechanism, fix $z_i=0$, $z_j=\infty$, and a third boundary puncture at a finite nonzero position. Equation~\eqref{eq:bulkzerolocus} removes the dependence of the bulk part of the Koba--Nielsen factor on all other finite boundary punctures, while the scalar kernel is already independent of the bulk coordinates. A simultaneous positive dilation $z_{a_\alpha}\mapsto\lambda z_{a_\alpha}$ and $z_{\bar a_\alpha}\mapsto\lambda z_{\bar a_\alpha}$ preserves the bulk domain and measure, and rescales this factor by $\lambda^\kappa$. Separating the common scale therefore gives
\begin{equation}
 \begin{gathered}
 \left.\bF_{n;m}^{\mathcal L_{\phi^3}}
 \right|_{\mathcal K_{\rm bulk}^{(i,j)}}
 =\mathcal H_{n;m}\int_0^\infty\frac{\dd\lambda}{\lambda}\lambda^\kappa,\\
 \kappa=\alpha'(Q^2-2p_i\cdot Q),\qquad
 Q=\sum_{\alpha=1}^m q_\alpha.
 \end{gathered}
 \label{eq:bulkzeroscale}
\end{equation}
Here $\mathcal H_{n;m}$ contains the boundary and relative-position integrations and is independent of $\lambda$. The scale integral has no common domain of convergence at its two endpoints. Under analytic continuation, the contributions from $(0,1)$ and $(1,\infty)$ are $1/\kappa$ and $-1/\kappa$, respectively, and cancel \cite{ArkaniHamed:2023zero,He:2026ngn}. The zero holds wherever the remaining integrations are regular. Appendix~\ref{app:bulkzero} gives the explicit change of variables and the full proof.

Nonzero mutual couplings \(q_\alpha\cdot q_\beta\) preserve the homogeneity under a common rescaling of all bulk punctures and therefore do not obstruct the bulk zero. Unlike the decoupling split, this zero does not require the insertions to decouple and is already present for a single insertion.

For gluons, also impose
\begin{equation}
 \epsilon_r\cdot q_\alpha=0
 \qquad
 r\notin\{i,j\},\quad \alpha\in[m].
 \label{eq:bulkzeropolarizations}
\end{equation}
This condition removes the bulk-coordinate terms from $V_r$ in Eq.~\eqref{eq:VWgeneral}, while $W_{rs}$ is already bulk independent. The full Wick kernel therefore becomes independent of the bulk coordinates, and the same scale argument yields
\begin{equation}
 \left.\bF_{n;m}^{\mathcal O_{\rm YM}}\right|_{\substack{\mathcal K_{\rm bulk}^{(i,j)}\\ \epsilon_r\cdot q_\alpha=0\ (\forall r,\alpha)}}=0
 \label{eq:bulkzeroYM}
\end{equation}
away from poles. The argument also applies to the mixed YMS kernel, since its flavor-trace Parke--Taylor factors depend only on boundary punctures.

The finite-$\alpha'$ zero descends to the normalized field-theory limit at regular kinematics. Whenever the merger decomposition~\eqref{eq:partitionformula} closes within the same operator family, as in the examples of Sec.~\ref{sec:contacts}, the single-insertion zero further implies that each partition contribution vanishes separately. The key observation is that the zero locus is preserved by merging insertions: for every block $B$,
\begin{equation}
 p_r\cdot Q_B=\sum_{\alpha\in B}p_r\cdot q_\alpha=0,
 \qquad r\notin\{i,j\}.
 \label{eq:bulkzeromerger}
\end{equation}
For polarized kernels, $\epsilon_r\cdot Q_B=0$ is inherited in the same way.


\section{Conclusion and discussion}
\label{sec:discussion}
In this paper, We have extended the disk construction of stringy form factors from a single operator insertion to multiple insertions. We derived a relation between the resulting stringy general form factors and open-string amplitudes and used it to evaluate the field-theory limit. Explicit examples show how the disk integral encodes both propagation between distinct insertions and local contact terms in which several insertions share a vertex. Set partitions organize these contributions, with each block specifying the insertion labels attached to a common vertex. Matching the resulting expressions to tree-level form factors then identifies the associated operators and fixes their relative weights. We also find that stringy general form factors exhibit the 2-split and the hidden zero, familiar from scattering amplitudes. In addition, we identify two new structures specific to form factors, which we call the decoupling split and the bulk zero.

An open question is how to determine the local operators and coefficients in Eq.~\eqref{eq:partitionformula} at arbitrary multiplicity. The formal source expansion in Eq.~\eqref{eq:sourceaction} offers an organizing picture, but does not by itself fix these data.

Another direction is to extend this construction to loop level and to superstrings. The annulus provides a starting point for the loop extension \cite{Stieberger:2021loop}, while superstring amplitudes with bulk insertions offer a route to the supersymmetric generalization \cite{Mafra:2013disk,BischofHaackStieberger:2023}. It would be interesting to determine how the field-theory limit, splits, and zeros extend to these settings.

Hidden zeros determine tree amplitudes in $\Tr(\phi^3)$ theory within an ordered local cubic ansatz with constant numerators, up to overall normalization \cite{Rodina:2024zero}. This result motivates a bootstrap of multi-insertion form factors using the splits and zeros found here. Combining their behavior on special kinematic loci with locality and factorization could constrain the contact contributions and enable the reconstruction of form factors at higher multiplicity.

\begin{acknowledgments}
 The work of Q.C. is supported by the Westlake Fellows Program at Westlake University. OpenAI's GPT-6 model, accessed through Codex, was used during derivations, computational checks, and manuscript preparation. The authors are responsible for the scientific content.
\end{acknowledgments}

\appendix

\section{Field-theory expansion and examples}
\label{app:wick-assembly}
This appendix derives the field-theory formula of Sec.~\ref{sec:field-theory} and supplies the conventions, normalization checks, and examples used in Sec.~\ref{sec:contacts}.

\subsection{Derivation of the field-theory expansion}
\label{app:monodromy-full}
The normalization in Eq.~\eqref{eq:normalizedexpansion} removes the largest pole of the complete ordered integrals, including the explicit $1/\alpha'$ factors in $W_{ij}$. Let $N=n+2m$ and let $t$ count the boundary scalar flavor traces, with $t=1$ for the scalar kernel and $t=0$ for the pure-gluon kernel. Together with the image pairs, the correlator has $c=t+m$ scalar cycles. In the logarithmic IBP reduction, connecting these cycles supplies $c-1$ powers of $\alpha'$ relative to a single-cycle integral, whose pole order is at most $N-3$ \cite{He:2019drm}. Counting the endpoint poles and Wick prefactors together therefore gives the generic pole order
\begin{equation}
 r_I=N-3-(c-1)=n+m-t-2.
\label{eq:sectorpoleorder}
\end{equation}
The field-theory prefactor is thus $(\alpha')^{r_I-m}=(\alpha')^{n-t-2}$.

With the normalized integrals regular at $\alpha'=0$, we can now derive the phase-moment formula~\eqref{eq:mthmoment}. Fix a complete kernel and suppress $I$ and the multiplicity on $\widehat A$. Write $\widehat A_w=\sum_{r\geq0}(\alpha')^r A_w^{(r)}$ and $\widehat\Phi_w=\Phi_{n,m}(w)/\alpha'$, using the invariants defined in Eq.~\eqref{eq:dimensionfulinvariants}. The phase is quadratic in ordering indicators. For disjoint nonempty words $W_1,\ldots,W_k$ covering the labels other than $1$, we first prove
\begin{equation}
\sum_{\sigma\in W_1\shuffle\cdots\shuffle W_k}\widehat A(1,\sigma)=O((\alpha')^{k-1}).
\label{eq:appshufflebound}
\end{equation}
The ordered integrals obey the trace representation
\begin{equation}
\sum_{\sigma\in S_{N-1}}\Tr(T_1T_\sigma)\widehat A(1,\sigma)=\sum_{\rho\in S_{N-2}}\Tr(T_1D_\rho)\widehat A(1,\rho,N).
\label{eq:apptraceidentity}
\end{equation}
Here $T_\sigma$ is the ordered matrix product, $D_\rho=D_{\rho,2}$, $D_{\rho,N}=T_N$, and, for $j=N-1,\ldots,2$,
\begin{equation}
\begin{aligned}
D_{\rho,j}&=T_{\rho_j}D_{\rho,j+1}-e^{-i\alpha'\gamma_{\rho,j}}D_{\rho,j+1}T_{\rho_j},\\
\gamma_{\rho,j}&=\pi\left(\hat{s}_{1,\rho_j}+\sum_{h=2}^{j-1}\hat{s}_{\rho_h,\rho_j}\right).
\end{aligned}
\label{eq:appDrecursion}
\end{equation}
Equation~\eqref{eq:apptraceidentity} follows from the left-nested deformed-commutator representation of Ref.~\cite[Sec.~2.2]{ChenDuMa:2012}, with $[X,Y]_q=XY-qYX$. Moving the commutators through the trace using $\Tr([X,Y]_qZ)=\Tr(X(YZ-qZY))$ gives Eq.~\eqref{eq:appDrecursion}. Each contour crossing supplies $e^{-i\pi s_{i,j}}$, while integer powers retain their signs in the rational correlator. The relation applies to the complete kernel. Gluon transversality and momentum conservation give $\sum_{j\ne i}\epsilon_i\cdot p_j=0$, canceling the noncovariant terms in $V_i$ and ensuring the required M\"obius weights. Individual gauge-fixed Wick monomials need not obey the relation. The contour identities are first derived with convergent endpoint regulators and then continued meromorphically to Eq.~\eqref{eq:generalimages}, without requiring an on-shell decomposition of the image momenta.

For each word $W_a=(i_{a,1},\ldots,i_{a,\ell_a})$, introduce an auxiliary space of dimension $\ell_a+1$ and form their tensor product. Let $T_{i_{a,j}}$ act as the matrix unit $E_{j,j+1}$ on the $a$th factor and as the identity on all others. Define $T_1$ to act as $E_{\ell_a+1,1}$ on every factor. Then $\Tr(T_1T_\sigma)$ selects precisely the required shuffles. In the recursion, the first matrix entering each new slot commutes with the inner expression and supplies $1-e^{-i\alpha'\gamma}=O(\alpha')$. There are $k-1$ such entries, proving Eq.~\eqref{eq:appshufflebound} by regularity of the normalized ordered integrals $\widehat A_w$.

The chambers are shuffles of the boundary word $(2,\ldots,n)$ and the $2m$ singleton images. A degree-$d$ indicator monomial $M_w$ fixes $d$ relative orderings. Adding these comparisons as directed edges to the boundary chain leaves at least $2m+1-d$ connected components. A directed cycle selects no chambers. Otherwise, for each allowed internal ordering of the components, the remaining freedom is a shuffle of their words. The selected chambers thus form a disjoint union of shuffles, and Eq.~\eqref{eq:appshufflebound} gives
\begin{equation}
\sum_{w\in\cP_{n,m}}M_w\widehat A_w=O((\alpha')^{2m-d}),\quad 0\leq d\leq2m.
\label{eq:appmonomialbound}
\end{equation}
Since $\widehat\Phi_w^p$ has degree at most $2p$, this gives
\begin{equation}
S_p:=\sum_w\widehat\Phi_w^p\widehat A_w=O((\alpha')^{2m-2p}),\quad 0\leq p\leq m.
\label{eq:appphasemomentbound}
\end{equation}
For $\widehat{\mathcal U}:=\sum_w e^{i\pi\alpha'\widehat\Phi_w}\widehat A_w$, the joint expansion is
\begin{equation}
[(\alpha')^q]\widehat{\mathcal U}=\sum_{p=0}^q\frac{(i\pi)^p}{p!}[(\alpha')^{q-p}]S_p.
\label{eq:appUcoefficient}
\end{equation}
Every term with $q<m$ vanishes, and only $p=m$ survives for $q=m$. Multiplication by $(i/2)^m$ proves Eq.~\eqref{eq:mthmoment}.

For a concrete example, set $m=2,n=2$ and $(a_1,\bar a_1,a_2,\bar a_2)=(3,4,5,6)$. In the scalar sector the ordered-integral correlator is
	\begin{equation}
		\frac{\PT(1,2)}{x_{34}^2x_{56}^2}
		=\PT(1,2)\PT(3,4)\PT(5,6).
		\label{eq:sixpointcorrelator}
	\end{equation}
Its leading scalar coefficient is $A_6^{\rm YMS}((1,2)|(3,4)|(5,6);\varnothing\Vert w)$. Equation~\eqref{eq:mthmoment} gives
	\begin{equation}
		\left.\bF_{2;2}^{(I)}\right|_{\rm FT}
		=\frac{\pi^2}{8}\sum_{\sigma\in S_5}
		\widehat\Phi_{2,2}(1,\sigma)^2 A_6^{(I,0)}(1,\sigma),
		\label{eq:sixpointmoment}
	\end{equation}
where $\sigma$ permutes $\{2,3,4,5,6\}$. The scalar traces remain $(1,2)\mid(3,4)\mid(5,6)$.

\subsection{Theory and diagrammatic conventions}
The comparison with tree-level operator insertions uses the following auxiliary theory and diagrammatic conventions. The auxiliary amplitudes are those of generalized Yang--Mills--scalar theory \cite[Eq.~(4.6)]{CachazoHeYuan:2015YMS},
\begin{align}
\mathcal L_{\rm YMS}={}&-\widetilde{\Tr}\left(\tfrac14F^{\mu\nu}F_{\mu\nu}
+\tfrac12D^\mu\phi^I D_\mu\phi^I\right)\notag\\
+\frac{g^2}{4}&\widetilde{\Tr}\sum_{I\ne J}[\phi^I,\phi^J]^2+\frac{\lambda}{3!}f_{IJK}f_{\tilde I\tilde J\tilde K}
\phi^{I\tilde I}\phi^{J\tilde J}\phi^{K\tilde K}.
\label{eq:LYMS}
\end{align}
Here $\widetilde{\Tr}$ denotes the gauge trace, and $I,J,K$ are flavor indices. Scalar traces specify the flavor ordering, while $w$ specifies the gauge ordering.

We strip color factors, couplings, and the common S-matrix phase. With outgoing momenta $k,l$ at a quadratic insertion, use
\begin{equation}
 \begin{gathered}
 V_\phi(k,l)=k\cdot l,\quad V_A^{\mu\nu}(k,l)=l^\mu k^\nu-(k\cdot l)\eta^{\mu\nu},\\
 D_\phi(k)=\frac1{k^2},\qquad D_A^{\mu\nu}(k)=-\frac{\eta^{\mu\nu}}{k^2}.
\end{gathered}
\end{equation}
The gluon propagator is in Feynman gauge; longitudinal contributions decouple since the insertion vertex is transverse. These conventions give $\epsilon_{1\mu}\epsilon_{2\nu}V_A^{\mu\nu}(p_1,p_2)=\Tr(f_1f_2)/2$, fixing the normalization of the operator denoted by $\Tr(F^2)$ in the main text. For the first exchange graph let $k=q_1-p_1=p_2-q_2$. Its scalar numerator is $(p_1\cdot k)(-p_2\cdot k)$, while contraction of the two gluon vertices with $-\eta_{\mu\nu}$ gives $q_1\cdot f_1f_2\cdot q_2$. Thus the graph values are
\begin{equation}
 \frac{(p_1\cdot q_1)(p_2\cdot q_2)}{(q_1-p_1)^2},\quad
 \frac{q_1\cdot f_1f_2\cdot q_2}{(q_1-p_1)^2}
 =-\frac{\operatorname{tr}(1,2)}{\hat{s}_{1,q_1}-q_1^2}.
\end{equation}
Interchanging the external labels gives the second graphs. A single quadratic insertion gives Eqs.~\eqref{eq:m2scalarcontact} and \eqref{eq:m2YMcontact} without a propagator. The local contact contribution in the disk decomposition is represented by that same single-insertion form factor with momentum $Q$.

\subsection{Two-gluon normalization and an exact six-point check}
\label{app:twogluonnorm}
The vertices above provide the reference normalization for the tree-level comparison. Matching them to the raw gluon disk kernel requires accounting for their different external-state normalizations. Relative to Ref.~\cite{HeTengZhang:2019string}, whose invariants are $s_{i,j}=\alpha'k_i\cdot k_j$, Eq.~\eqref{eq:VWgeneral} uses $k_i=\sqrt2p_i$ and $E_i=\sqrt2\epsilon_i$ for boundary gluons. The image invariants are imposed directly as in Eq.~\eqref{eq:generalimages}. Equation~\eqref{eq:YMSleading} uses this worldsheet convention; the tree form factors use the vertices stated above.

In the cyclic amplitude convention, the scalar one-insertion coefficient is $-\pi F_{\mathcal O;2}$, and the two-insertion coefficient is the $+\pi^2$ combination in Eq.~\eqref{eq:m2scalardecomposition}. Comparing the directed disk integral with this convention requires consistent orientations of the contours, Faddeev--Popov factors, and ordered integrals. A coordinate Jacobian or a reduced Pfaffian with unspecified orientation does not determine the common phase by itself. The relative gluon factor below is independent of this phase.

A direct check is obtained by taking both gluon polarizations orthogonal to all external and insertion momenta, with $e=\epsilon_1\cdot\epsilon_2$. Then $V_1=V_2=0$ and
\begin{equation}
\begin{gathered}
\mathcal I_{2;2}^{\mathcal O_{\rm YM}}
=-\frac{2e}{\alpha'}\mathcal I_{2;2}^{\mathcal L_{\phi^3}},\\
\left.\bF_{2;2}^{\mathcal O_{\rm YM}}\right|_{\rm FT}
=-2e\left.\bF_{2;2}^{\mathcal L_{\phi^3}}\right|_{\rm FT}.
\end{gathered}
\label{eq:transversenorm}
\end{equation}
The tree expressions instead obey $F_{\rm sep}^{\rm YM}=-eF_{\rm sep}^{\phi}$ and $F_{\rm single}^{\rm YM}=-eF_{\rm single}^{\phi}$. This fixes the relative factor of two in Eq.~\eqref{eq:m2YMdecomposition}.

The remaining tensor structure can be checked at the nonsingular invariant point
\begin{equation}
\begin{gathered}
p_1\cdot p_2=7,\quad p_1\cdot q_1=2,\quad p_2\cdot q_1=3,\quad q_1^2=11,\\
 q_2=p_1+p_2-q_1.
\end{gathered}
\label{eq:sixpointnormpoint}
\end{equation}
Define $a=\epsilon_1\cdot p_2$, $b=\epsilon_2\cdot p_1$, $c=\epsilon_1\cdot q_1$, $d=\epsilon_2\cdot q_1$, and $e=\epsilon_1\cdot\epsilon_2$. Evaluating all $120$ chambers with \texttt{Amp.m} gives the exact moments
\begin{equation}
\begin{aligned}
\sum_w\widehat\Phi_w^2 A_w^{(\phi,0)}&=\frac{624}{7},\\
\sum_w\widehat\Phi_w^2 A_w^{({\rm YM},0)}
&=\frac{1056ab-576bc-384ad}{35}+\frac{192cd}{5}-\frac{1248e}{7}.
\end{aligned}
\label{eq:sixpointnormmoments}
\end{equation}
The corresponding tree sums are
\begin{equation}
\begin{aligned}
F_{\rm sep}^{\phi}+F_{\rm single}^{\phi}=&\frac{78}{7},\\
F_{\rm sep}^{\rm YM}+F_{\rm single}^{\rm YM}=&\frac{66ab-36bc-24ad}{35}+\frac{12cd}{5}-\frac{78e}{7}.
\end{aligned}
\label{eq:sixpointnormtrees}
\end{equation}
Multiplying the moments by $\pi^2/8$ verifies Eqs.~\eqref{eq:m2scalardecomposition} and \eqref{eq:m2YMdecomposition} for every polarization coefficient at this point. When comparing reduced Pfaffians to the raw correlator, the sign $(-1)^{N(N-1)/2}$ in Ref.~\cite[Appendix~F]{HeTengZhang:2019string} must be retained; it is positive at four points and negative at six points.

\subsection{Scalar examples with multiple insertions}
\label{app:n3m2}

With the normalization fixed in the minimal two-insertion example, we turn to scalar configurations that introduce cubic interactions and partial mergers. The first example has $n=3$ and $m=2$. We use the quadratic insertion and propagator conventions above, and set both the ordinary cubic scalar vertex and the cubic operator vertex to unity. Define the dimensionful quantities
\begin{equation}
\begin{gathered}
 Q=q_1+q_2=p_1+p_2+p_3,\\ x_{i\alpha}=p_i\cdot q_\alpha,\quad
 d_{i\alpha}=(q_\alpha-p_i)^2=q_\alpha^2-2x_{i\alpha},\\
 S_i=(Q-p_i)^2=\hat{s}_{j,k},\quad\{i,j,k\}=\{1,2,3\}.
\end{gathered}
\end{equation}
The separated-insertion form factor contains eighteen tree diagrams: six with one insertion on an external edge and the other at the cubic vertex, six with insertions on different edges, and six with both insertions on the same edge in either order. Their sum is
\begin{equation}
\begin{gathered}
 F_{\mathcal O,\mathcal O;3}
 =\sum_{i=1}^{3}\sum_{\alpha=1}^{2}
       \frac{x_{i\alpha}}{d_{i\alpha}}
 +\sum_{\substack{i,j=1\\i\ne j}}^{3}
       \frac{x_{i1}x_{j2}}{d_{i1}d_{j2}}\\+\sum_{i=1}^{3}\sum_{\alpha=1}^{2}
       \frac{x_{i\alpha}
       (q_{3-\alpha}^{2}-d_{i\alpha}-S_i)}
       {2d_{i\alpha}S_i}.
\end{gathered}
\label{eq:n3m2separated}
\end{equation}
This expression agrees algebraically with direct tree enumeration after imposing momentum conservation. The merged contribution is the single-insertion form factor
\begin{equation}
 F_{\mathcal O;3}(Q)=1+\sum_{i=1}^{3}
       \frac{x_{i1}+x_{i2}}{S_i}.
\label{eq:n3m2contact}
\end{equation}
The merged operator has both quadratic and cubic vertices. Its three-scalar form factor therefore contains a local cubic contact and diagrams with a propagator attached to the quadratic vertex. Comparing the seven-point YMS phase moment from \texttt{Amp.m}, with the assignment~\eqref{eq:generalimages} and normalization above, to these tree expressions gives
\begin{equation}
 \frac1{\pi^2}\left.\bF_{3;2}\right|_{\rm FT}
 =F_{\mathcal O,\mathcal O;3}+F_{\mathcal O;3}(Q).
\label{eq:n3m2identity}
\end{equation}

The two-insertion example contains separate insertions and a single merged contribution. With three insertions, two operators may also merge while the third remains separate. This additional structure is already present for three $\mathcal L_{\phi^3}$ insertions and two external scalars, with $q_1+q_2+q_3=p_1+p_2$. The field-theory expansion contains the five partitions of three insertion labels:
	\begin{equation}
\begin{gathered}
			-\frac{1}{\pi^3}\left.\bF_{2;3}^{\mathcal L_{\phi^3}}\right|_{\rm FT}
			=F_{1|2|3;2}
			+F_{(12)|3;2}
			\\+F_{(23)|1;2} 
			+F_{(31)|2;2}+F_{(123);2}.
		\end{gathered}
\label{eq:m3explicitdecomposition}
	\end{equation}
	\begin{equation}
\begin{gathered}
			F_{1|2|3;2}
			=\sum_{\sigma\in S_3}
			\frac{\hat{s}_{1,q_{\sigma_1}}\hat{s}_{2,q_{\sigma_2}}}
			{8(q_{\sigma_1}-p_1)^2(q_{\sigma_2}-p_2)^2} 
			\\{}\times\left[q_{\sigma_3}^2
			-(q_{\sigma_1}-p_1)^2-(q_{\sigma_2}-p_2)^2\right].
		\end{gathered}
\label{eq:m3separated}
	\end{equation}
Here $1|2|3$ denotes three separate insertions. In $(12)|3$, the first two operators merge into one insertion carrying $q_1+q_2$, while the third carries $q_3$. Writing $Q_{12}=q_1+q_2$ and $\hat{s}_{i,Q_{12}}=\hat{s}_{i,q_1}+\hat{s}_{i,q_2}$, Eq.~\eqref{eq:m2scalarcontact} gives
	\begin{equation}
\begin{aligned}
&F_{(12)|3;2}=F_{\mathcal O,\mathcal O;2}(Q_{12},q_3)\\
&=\frac{\hat{s}_{1,Q_{12}}(\hat{s}_{1,2}-\hat{s}_{2,Q_{12}})}{4(Q_{12}^2-\hat{s}_{1,Q_{12}})}
+\frac{(\hat{s}_{1,2}-\hat{s}_{1,Q_{12}})\hat{s}_{2,Q_{12}}}{4(Q_{12}^2-\hat{s}_{2,Q_{12}})}.
\end{aligned}
\label{eq:m3mergedpair}
\end{equation}
The other two pair-merger terms follow by cyclically permuting $(q_1,q_2,q_3)$. When all three operators merge, the result is
	\begin{equation}
		F_{(123);2}=F_{\mathcal O;2}(q_1+q_2+q_3)=\frac{\hat{s}_{1,2}}2.
		\label{eq:m3mergedall}
	\end{equation}
As in Eq.~\eqref{eq:m2scalarcontact}, the invariants in this example are dimensionful: $\hat{s}_{i,q_\alpha}=2p_i\cdot q_\alpha$ and $\hat{s}_{1,2}=2p_1\cdot p_2$.

	\section{Proof of the split identities}
	\label{app:splitlemma-full}

We return from the field-theory expansion to the finite-$\alpha'$ split identities of Sec.~\ref{sec:splits-zeros}. Both scalar splits follow directly from the disk representation by removing the couplings between two groups of punctures. We first establish their common structure for the scalar kernel, then identify the additional conditions needed to separate the gluon Wick contractions.

	\subsection{Universal split lemma}
	
The two splits can be treated together by denoting their non-shared label sets by $L$ and $R$. For the same-side split they are
	\begin{equation}
		(L,R)_{\rm same}=\left(A,
		B\cup\mathsf Q_{m}\right),
		\label{eq:appLRsame}
	\end{equation}
whereas for the decoupling split they are
	\begin{equation}
		(L,R)_{\rm dec}=\left(
		A\cup\mathsf Q_{M_A},
		B\cup\mathsf Q_{M_B}\right).
		\label{eq:appLRdec}
	\end{equation}
Both cases obey the condition on each image label
	\begin{equation}
		s_{a,b}=0,\qquad a\in L,\quad b\in R.
		\label{eq:appstrongsplit}
	\end{equation}
	
Fix $z_i=0$, $z_j=1$, and $z_k=\infty$.  The logarithm of the doubled Koba--Nielsen factor is
	\begin{equation}
		\mathcal S_{n;m}=\sum_{X<Y}s_{X,Y}\log z_{X,Y}.
	\end{equation}
Every term joining $L$ and $R$ vanishes on Eq.~\eqref{eq:appstrongsplit}; therefore
	\begin{equation}
		\mathcal S_{n;m}=\mathcal S_L+\mathcal S_R,
		\qquad \KN_{n;m}=\KN_L\KN_R.
		\label{eq:appKNsplit}
	\end{equation}
Self-image invariants stay within one side and cause no obstruction.  The same argument applied to the continuation phase \eqref{eq:generalphase} gives, chamber by chamber,
	\begin{equation}
		\Phi_{n,m}=\Phi_L+\Phi_R,
		\qquad e^{i\pi\Phi_{n,m}}=e^{i\pi\Phi_L}e^{i\pi\Phi_R}.
		\label{eq:appphasesplit}
	\end{equation}
	
After fixing the common punctures, the boundary factors separate for a compatible cyclic ordering. The $\PT$ symbols below denote gauge-fixed chains, including the inherited Faddeev--Popov and infinity factors:
	\begin{equation}
		\PT(1,\ldots,n)=
		\PT(i,A,j,\kappa_L)\PT(j,B,i,\kappa_R).
		\label{eq:appPTsplit}
	\end{equation}
The integration domain is a Cartesian product.  Its measure is
	\begin{equation}
		\dd\mu_{n;m}=\begin{cases}
			\dd\mu^L_{|A|+3;0}\dd\mu^R_{|B|+3;m},
			&\text{same side},\\
			\dd\mu^L_{|A|+3;|M_A|}
			\dd\mu^R_{|B|+3;|M_B|},
			&\text{decoupling}.
		\end{cases}
		\label{eq:appmeasuresplit}
	\end{equation}
In the all-outgoing convention, each image pair contributes $-q_\alpha$ to a linear momentum sum. This is bookkeeping for the analytic invariant assignment, not an on-shell momentum decomposition. The auxiliary momenta are
	\begin{equation}
		p_{\kappa_L}=-p_i-p_j-\sum_{a\in L}p_a,
		\quad p_{\kappa_R}=-p_i-p_j-\sum_{b\in R}p_b.
		\label{eq:appkappa}
	\end{equation}
Combining Eqs.~\eqref{eq:appKNsplit}--\eqref{eq:appmeasuresplit} proves Eqs.~\eqref{eq:samesidesplit} and \eqref{eq:sepsplit} for the scalar kernel.  Since no pole limit was taken, these are identities of disk integrals restricted to the split subspace, rather than factorization residues.
	
For mixed kernels, compatibility is imposed trace by trace: no adjacent pair in any cyclic word $\rho^{(a)}$, including its closing pair, may join non-shared scalars on opposite sides. Otherwise the corresponding zero need not hold. At four points, take the shared labels to be $1,3$ and the non-shared sets to be $A=\{2\}$ and $B=\{4\}$. The trace $(2,4)$ crosses the split. In the stripped conventions of Eq.~\eqref{eq:LYMS}, $A_4^{\rm YMS}((1,3)|(2,4);\varnothing\Vert1,2,3,4)=-1$, so it remains nonzero on $s_{13}=s_{24}=0$.

For a compatible ordering, the split also allows a hidden zero of one current to become a zero of the full integral, provided the other current remains regular. As a concrete scalar example, choose $(i,j,k)=(1,3,5)$, $A=\{2\}$, and $B=\{4\}$. The scalar split conditions together with the inherited-zero condition are
\begin{equation}
s_{2,4}=s_{2,5}=s_{2,a_\alpha}=s_{2,\bar a_\alpha}=0,
\label{eq:n5zero}
\end{equation}
for every $\alpha$. Here $s_{2,5}=s_{2,\kappa_A}$ puts the four-point current on its zero. For gluons also impose Eq.~\eqref{eq:splitpolarizations} and $\epsilon_2\cdot p_5=0$.

    \subsection{Polarized kernels}
The preceding argument separates the measure, integration domain, and scalar kernel. Extending the split identities to gluons also requires the Wick contractions to separate. We assign the shared polarizations to the right factor, so the left current has gluons in $A$ and scalar legs $i,j,\kappa_A$; the right factor is contracted with $\epsilon_k$ \cite{Cao:2024universal,Cao:2025ff}.

Consider first the same-side split, with all image pairs in $R$ and $L=A$. The elementary contractions are
    \begin{equation}
        V_a=\sum_{X\ne a}\frac{2\epsilon_a\cdot p_X}{z_{a,X}},
        \qquad W_{a,b'}=\frac{2\epsilon_a\cdot\epsilon_{b'}}
        {\alpha' z_{a,b'}^2}.
        \label{eq:appVW}
    \end{equation}
Here $X$ runs over all boundary and image labels. Keeping $z_k$ finite makes the shared-leg contributions explicit. Define
    \begin{equation}
\begin{gathered}
        V_a^L:=\sum_{\substack{X\in A\cup\{i,j,k\}\\X\ne a}}
        \frac{2\epsilon_a\cdot p_X}{z_{a,X}},\qquad a\in A,\\
        V_{b'}^R:=\sum_{\substack{X\in R\cup\{i,j,k\}\\X\ne b'}}
        \frac{2\epsilon_{b'}\cdot p_X}{z_{b',X}},
        \quad b'\in B\cup\{i,j,k\}.
    \end{gathered}
\label{eq:appVWrestricted}
    \end{equation}
The superscripts specify the momentum labels retained in each contraction. The right contractions are defined for $b'\in B\cup\{i,j,k\}$, so the shared gluon polarizations belong to the right factor.

For $a\in A$ and $b'\in B\cup\{i,j,k\}$, Eq.~\eqref{eq:splitpolarizations} implies
    \begin{equation}
\begin{aligned}
        V_a-V_a^L=&\sum_{X\in R}\frac{2\epsilon_a\cdot p_X}{z_{a,X}}=0,\\
         V_{b'}-V_{b'}^R=&\sum_{a\in A}\frac{2\epsilon_{b'}\cdot p_a}{z_{b',a}}=0, \quad W_{a,b'}=0.
        \end{aligned}
\label{eq:appVWsplit}
    \end{equation}
The first equality uses $\epsilon_a\cdot p_b=0$ and $\epsilon_a\cdot q_\alpha=0$ for every image pair. The second equality uses $p_a\cdot\epsilon_{b'}=0$, including $b'=i,j,k$. The condition on $W_{a,b'}$ forbids every Wick pair joining a polarization in $A$ to a polarization in $B\cup\{i,j,k\}$. Thus each surviving matching is uniquely a matching on $A$ together with one on $B\cup\{i,j,k\}$, and the full matching sum factorizes.

Now fix $z_i=0,z_j=1$ and take $z_k\to\infty$ with the usual vertex and gauge-fixing factors. For $a\in A$, the term $2\epsilon_a\cdot p_k/z_{a,k}$ tends to zero, while the terms involving $p_i,p_j$ remain. The polarization $\epsilon_k$ belongs entirely to the right matching sum: its contractions with $A$ vanish already before the infinity limit. Its vertex must be taken to infinity with its conformal prefactor, rather than discarded as a vanishing $V_k$.

The Wick kernel now factorizes together with the measure and Koba--Nielsen factor. The left current carries gluons in $A$ and scalar legs $i,j,\kappa_A$, while the right current retains the shared polarizations and is contracted with $\epsilon_k$ \cite{Cao:2024universal,Cao:2025ff}. The scalar legs carry momentum but no polarization. The split allows nonzero $\epsilon_a\cdot p_i$, $\epsilon_a\cdot p_j$, and $\epsilon_a\cdot p_k$; the inherited hidden zero additionally sets the last contraction to zero. This proves Eq.~\eqref{eq:samesidesplitYM} for every $m$, with all image pairs on the right.

For insertions on both sides, set $L=A\cup\mathsf Q_{M_A}$ and $R=B\cup\mathsf Q_{M_B}$, with all shared polarizations retained on the right. Equations~\eqref{eq:oppositeside} and \eqref{eq:decpolarizations} now remove contractions between boundary polarizations and the opposite bulk group, including those of the shared legs. Each operator is represented by both image labels. Before taking $z_k$ to infinity, define
    \begin{equation}
\begin{gathered}
      V_a^L:=\sum_{\substack{X\in L\cup\{i,j,k\}\\X\ne a}}
      \frac{2\epsilon_a\cdot p_X}{z_{a,X}},\qquad a\in A,\\
      V_{b'}^R:=\sum_{\substack{X\in R\cup\{i,j,k\}\\X\ne b'}}
      \frac{2\epsilon_{b'}\cdot p_X}{z_{b',X}},
      \quad b'\in B\cup\{i,j,k\}.
    \end{gathered}
\label{eq:appdecV}
    \end{equation}
Unlike the same-side case, both expressions contain bulk terms. The differences from the full contractions are
    \begin{equation}
    \begin{aligned}
      V_a-V_a^L
      ={}&\sum_{b\in B}\frac{2\epsilon_a\cdot p_b}{z_{a,b}}-\sum_{\beta\in M_B}\epsilon_a\cdot q_\beta
      \left(\frac1{z_{a,a_\beta}}+
      \frac1{z_{a,\bar a_\beta}}\right),\\
      V_{b'}-V_{b'}^R
      ={}&\sum_{a\in A}\frac{2\epsilon_{b'}\cdot p_a}{z_{b',a}}-\sum_{\alpha\in M_A}\epsilon_{b'}\cdot q_\alpha
      \left(\frac1{z_{b',a_\alpha}}+
      \frac1{z_{b',\bar a_\alpha}}\right).
    \end{aligned}
    \label{eq:appdecVdifference}
    \end{equation}
Every term vanishes under Eq.~\eqref{eq:decpolarizations}. In particular, $\epsilon_{b'}\cdot q_\alpha=0$ includes $b'=i,j,k$: it removes the shared polarizations' dependence on the left bulk coordinates. The momentum conditions alone do not remove these terms. Also $W_{a,b'}=0$ for all $a\in A$ and $b'\in B\cup\{i,j,k\}$.

Consequently every surviving Wick matching is a disjoint union of a matching on $A$ and one on $B\cup\{i,j,k\}$. Replacing the singleton contractions by $V_a^L$ and $V_{b'}^R$ therefore makes the full matching sum a product of two sums. After fixing $z_i=0,z_j=1,z_k=\infty$, the left sum depends only on left integration coordinates and the right sum only on right ones. The $k$ vertex, including its conformal prefactor, belongs to the right factor; its cross contractions vanish before the limit. The shared momenta $p_i,p_j$ still enter both factors, whereas the shared polarizations occur only on the right.

Combining this kernel decomposition with Eqs.~\eqref{eq:appKNsplit} and \eqref{eq:appmeasuresplit} gives two form-factor currents. For the scalar kernel, these are the $\bF^{\mathcal L_{\phi^3}}$ currents in Eq.~\eqref{eq:sepsplit}. For the gluon kernel, the left factor is $\bF^{\rm mixed}$, with gluons in $A$, scalar legs $i,j,\kappa_A$, and insertions in $M_A$; the right auxiliary index is contracted with $\epsilon_k$. This proves Eq.~\eqref{eq:sepsplitYM}. When $M_A=\varnothing$, the left current contains no bulk insertions and reduces to the amplitude current $\bJ$, recovering the same-side split.

	\section{Proof of the bulk zero}
	\label{app:bulkzero}
	
The bulk zero of Sec.~\ref{sec:bulk-zeros} follows from the common scale of the bulk punctures. On Eq.~\eqref{eq:bulkzerolocus}, a simultaneous dilation separates a scaleless integral while the relative positions remain coupled. The mutual bulk couplings therefore need not vanish, and we retain them throughout the proof.
	
For $n\geq3$, fix $z_i=0$, $z_j=\infty$, and a third boundary position to a finite value. Keep the bulk coordinates $z_{a_\alpha},z_{\bar a_\alpha}$ of Eq.~\eqref{eq:diskdef}. On Eq.~\eqref{eq:bulkzerolocus}, the bulk-dependent part of the Koba--Nielsen factor is
	\begin{equation}
\begin{gathered}
			\prod_{\alpha=1}^{m}
			(z_{a_\alpha}z_{\bar a_\alpha})^{-\alpha'p_i\cdot q_\alpha}
			(z_{a_\alpha,\bar a_\alpha})^{\alpha'q_\alpha^2} 
			\\{}\times\prod_{\alpha<\beta}
			\Big[z_{a_\alpha,a_\beta}z_{a_\alpha,\bar a_\beta}
			z_{\bar a_\alpha,a_\beta}z_{\bar a_\alpha,\bar a_\beta}
			\Big]^{\frac{\alpha'}2q_\alpha\cdot q_\beta}.
		\end{gathered}
\label{eq:appbulkfactor}
	\end{equation}
Each complex power retains the branch inherited from the disk integral; the brackets abbreviate the product of the four powers. Under the simultaneous positive dilation $z_{a_\alpha}\mapsto\lambda z_{a_\alpha}$, $z_{\bar a_\alpha}\mapsto\lambda z_{\bar a_\alpha}$ for all $\alpha$, each measure $\dd^2z_{a_\alpha}/(z_{a_\alpha,\bar a_\alpha})^2$ is invariant. The factor in Eq.~\eqref{eq:appbulkfactor} acquires $\lambda^\kappa$, where $Q=\sum_\alpha q_\alpha$ and
	\begin{equation}
		\kappa=\alpha'\left[
		\sum_\alpha q_\alpha^2
		+2\sum_{\alpha<\beta}q_\alpha\cdot q_\beta
		-2p_i\cdot\sum_\alpha q_\alpha\right].
		\label{eq:appbulkhomogeneity}
	\end{equation}
This dilation preserves the bulk domain and the inherited branch.
	
\subsection{Separating the common scale}
The common scale separates explicitly upon choosing
\begin{equation}
\begin{gathered}
 \lambda=\operatorname{Im}z_{a_1}>0,\quad z_{a_1}=\lambda(u_1+i),\\
 z_{a_\alpha}=\lambda w_\alpha\quad(\alpha\ge2), \quad u_1\in\mathbb R,\quad w_\alpha\in\mathbb H_+.
\end{gathered}
\label{eq:bulkscalevariables}
\end{equation}
The Jacobian gives
\begin{equation}
\begin{gathered}
 \prod_{\alpha=1}^m\frac{\dd^2z_{a_\alpha}}{(z_{a_\alpha}-\bar z_{a_\alpha})^2}
 =\frac{\dd\lambda}{\lambda}\,\dd\nu,\\
 \dd\nu=\frac{\dd u_1}{(2i)^2}
 \prod_{\alpha=2}^m\frac{\dd^2w_\alpha}{(w_\alpha-\bar w_\alpha)^2}.
\end{gathered}
\label{eq:bulkscalemeasure}
\end{equation}
All remaining variables have domains independent of $\lambda$. By Eq.~\eqref{eq:appbulkhomogeneity}, the only scale dependence is $\lambda^\kappa$, with $\kappa=\alpha'(Q^2-2p_i\cdot Q)$. Thus
\begin{equation}
 \left.\bF_{n;m}^{\mathcal L_{\phi^3}}\right|_{\mathcal K_{\rm bulk}^{(i,j)}}
 =\mathcal H_{n;m}
 \int_0^\infty\frac{\dd\lambda}{\lambda}\lambda^\kappa,
 \label{eq:bulkscaleless}
\end{equation}
where $\mathcal H_{n;m}$ contains the boundary and relative-position integrations. Following the scaleless-integral argument of Ref.~\cite[Sec.~6.1]{ArkaniHamed:2023zero}, split the scale integral at $\lambda=1$:
\begin{equation}
 \left[\int_0^1\dd\lambda\,\lambda^{\kappa-1}\right]_{\rm AC}
 +\left[\int_1^\infty\dd\lambda\,\lambda^{\kappa-1}\right]_{\rm AC}
 =\frac1\kappa-\frac1\kappa=0.
\label{eq:bulkscaleAC}
\end{equation}
The two ranges converge for $\operatorname{Re}\kappa>0$ and $\operatorname{Re}\kappa<0$, respectively. Their meromorphic continuations cancel, so the full integral vanishes wherever the remaining integrations are regular.

The same scale separation extends to polarized kernels under Eq.~\eqref{eq:bulkzeropolarizations}. This condition removes all bulk-coordinate dependence from $V_r$, while $W_{rs}$ is already bulk independent. The scaleless-integral argument therefore applies to the full YM/YMS kernel away from poles, without requiring the mutual bulk couplings to vanish, proving Eq.~\eqref{eq:bulkzeroYM}.

\subsection{A coupled two-insertion example}
A concrete example with $n=3,m=2$ shows that the bulk zero does not require $q_1\cdot q_2=0$. Choose $i=1,j=2$ and set $p_3\cdot q_1=p_3\cdot q_2=0$. A consistent analytic kinematic family is specified by
\begin{equation}
 \begin{gathered}
 p_1\cdot p_2=u,\quad p_1\cdot p_3=v,\quad p_2\cdot p_3=-v,\\
 p_1\cdot q_1=x,\quad p_2\cdot q_1=y,\quad q_1^2=t,
 \\ q_2=p_1+p_2+p_3-q_1.
\end{gathered}
\end{equation}
Here $Q^2=2u$, $q_1\cdot q_2=x+y-t$, $q_2^2=2u+t-2x-2y$, and $\kappa=-2\alpha'v$. At $(u,v,x,y,t)=(7,2,2,3,11)$, one has $q_1\cdot q_2=-6$ and $\kappa=-4\alpha'$, so the bulk punctures remain coupled.

The field-theory expressions provide a direct check at this point. In the notation of Eq.~\eqref{eq:n3m2separated},
\begin{equation}
 \begin{gathered}
 (x_{11},x_{21},x_{31})=(2,3,0),\\
 (x_{12},x_{22},x_{32})=(7,2,0),\\
 (S_1,S_2,S_3)=(-4,4,14),
\end{gathered}
\end{equation}
while $(d_{11},d_{21},d_{31})=(7,5,11)$ and $(d_{12},d_{22},d_{32})=(1,11,15)$, so all propagator denominators are nonzero. The three sums in Eq.~\eqref{eq:n3m2separated} give
\begin{equation}
 \begin{gathered}
 F_{\mathcal O,\mathcal O;3}
 =\frac{3106}{385}+\frac{1637}{385}-\frac{4743}{385}=0,\\
 F_{\mathcal O;3}(Q)=1-\frac94+\frac54=0.
\end{gathered}
\end{equation}
Thus both the separated and merged contributions vanish, in agreement with the bulk zero and Eq.~\eqref{eq:n3m2identity}.

\bibliography{two_closed_stringy_form_factors}

\end{document}